# THERMAL AND THERMOELECTRIC TRANSPORT IN NANOSTRUCTURES AND LOW-DIMENSIONAL SYSTEMS

Li Shi

Department of Mechanical Engineering and Texas Materials Institute, The University of Texas at Austin, Austin, TX 78712

lishi@mail.utexas.edu

Abstract

Significant progress has been made in recent studies of thermal and thermoelectric transport phenomena in nanostructures and low-dimensional systems. This article reviews several intriguing quantum and classical size effects on thermal and thermoelectric properties that have been predicted by theoretical calculations or observed in experiments. Attention is focused on the Casimir limit in phonon boundary scattering and the effect of phonon confinement on the lattice thermal conductivity of semiconductor nanowires (NWs) and nanomeshes; the effects of thickness, lateral size, and interface interaction on the lattice thermal conductivity of carbon nanotubes (CNTs) and graphene; and the phonon-drag thermopower and quantum size effects on the thermoelectric power factor in semiconductor NWs. Further experimental and theoretical investigations are suggested for better understanding of some of these nanoscale transport phenomena.

## INTRODUCTION

In recent years, one focus of thermophysical engineering research has been to address the increasingly severe heat dissipation problem in microelectronic and other functional devices. As the critical feature size of electronic devices is reduced to increase the packing density and to reduce cost, the power density is increased and can exceed 1000 W/cm$^2$ at local hot spots [1]. Understanding thermal transport processes around these hot spots and the size effects in the thermal properties of materials has been a topic of both scientific interest and practical importance [2]. In addition, the increasing awareness of the global challenge in energy and sustainability has recently rekindled the interest in efficient and environmentally friendly thermal energy conversion and storage technologies, including solid-state thermoelectric (TE) power generation and refrigeration based on the Seebeck and Peltier effects [3]. Various size effects have been investigated as possible new approaches to enhancing the transport properties of materials for these applications. These properties include the thermal conductivity ($\kappa$), the thermopower or Seebeck coefficient ($S$), and the electrical conductivity ($\sigma$).

The room-temperature $\kappa$ of dense bulk inorganic solids span the range between about 1 W m$^{-1}$ K$^{-1}$ in amorphous materials such as silica to 2230 W m$^{-1}$ K$^{-1}$ reported for natural diamond, although  a higher value of 3320 W m$^{-1}$ K$^{-1}$ has been suggested for isotopically enriched C-12 diamond [4, 5]. High-$\kappa$ materials are used for electronic cooling and thermal energy storage. In comparison, low-$\kappa$ materials, such as the recently reported disordered, layered thin films [6] of a ultralow $\kappa$ value not much higher than that of air, are desirable for thermal insulation and TE waste heat recovery and refrigeration. The energy efficiency of TE devices is governed by the materials figure of merit, which is defined as

$$Z \equiv S^2 \sigma / \kappa, \qquad (1)$$

Figure 1 shows that the dimensionless $ZT$ product reaches the peak value close to unity within a narrow temperature ($T$) range near room, intermediate, and high temperature, respectively, for $Bi_2Te_3$, PbTe, and SiGe alloys [7-13]. For TE conversion to become sufficiently efficient and economical for vehicle waste heat recovery, the $ZT$ value needs to be increased above 2 over a large temperature range. A larger average $ZT$ value may be needed to make solar-thermal-electricity conversion and TE refrigeration a commercially viable option beyond the current niche markets including cooling car seats and optoelectronic devices [14-16]. Although there is no established theoretical limit in $ZT$, increasing $ZT$ to above unity has proven difficult because the interdependence of the three properties. For example, increasing $\sigma$ by increasing the charge carrier concentration can lead to a decrease in $S$ and an increase of the electronic contribution ($\kappa_e$) to $\kappa$ [17]. In addition, because traditional TE materials contain elements such as Te and Ge that are not abundant on earth and become increasingly costly due to their use in other technologies, alternative high-$ZT$ materials containing only earth-abundant elements need to be discovered before TE conversion can play a role in addressing the worldwide challenge in energy and sustainability.

Considerable advancements have been made in controlling materials structure down to the nanometer scale. When the characteristic dimension of the materials structure becomes comparable to the wavelength or mean free path of electrons and phonons that are the energy carriers in solids, quantum confinement and classical interface scattering effects can emerge and modify the transport properties.

The de Broglie wavelength of an electron is

$$\lambda_e = 2\pi\hbar / \sqrt{2m^*E} \quad ,\tag{2}$$

where $\hbar$ is the reduced Planck constant, $m^*$ is the electron effective mass, and $E$ is the electron energy. When $E$ measured from the band edge is of the order of the thermal energy $k_B T$, where $k_B$ is the Boltzmann constant, $\lambda_e$ can be as long as tens of nanometers in Bi, $Bi_{0.95}Sb_{0.5}$ and InSb that all have a very small $m^*$.

In comparison, the spectral energy density of phonons at low temperatures peaks at a wavelength ($\lambda_{p,0}$) given as

$$\lambda_{p,0} T \approx 2\pi \hbar v_s / x k_B, \qquad (3)$$

where $x$ takes a value close to 5, 4, and 3 for three-, two-, and one-dimensional (3D, 2D, and 1D) systems. For a typical sound velocity $v_s$ of the order of $10^4$ m/s, $\lambda_{p,0} T \approx 100$ nm. This expression breaks down at high temperature as the obtained $\lambda_{p,0}$ value becomes smaller than the minimum allowable wavelength $\lambda_{\min} = 2a$, where $a$ is the lattice constant. In this case, the spectral energy density peaks at $\lambda_{\min}$, corresponding to the edge of the Brillouin zone of the reciprocal lattice. Hence, except at very low temperatures or in crystals with a very large $a$, the spectral phonon density near room temperature peaks at a wavelength of the order of 1 nm. However, because of the lack of umklapp scattering for long-wavelength phonons, the spectral phonon thermal conductivity can peak at a longer wavelength than $\lambda_{p,0}$ or $\lambda_{\min}$.

The mean free path of electrons and phonons can range from as short as close to the atomic spacing to as long as more than microns in high quality crystals or at low temperatures. A long electron mean free path ($l_e$) is usually found in small-$m^*$ semiconductors or semimetals, such as InSb, Bi, and graphene and its derivatives including carbon nanotubes (CNTs). Here, the term *nanostructures* is broadly employed for those with a characteristic size in the sub-100 nm scale, which can be comparable to the mean free paths of the energy carriers. In comparison, the term *low-dimensional (low-D) systems* is used exclusively for those with the characteristic size of one or more dimensions being comparable to or smaller than the wavelength of electrons or phonons.

A number of quantum and classical size effects on thermal and TE properties in low-D systems and nanostructures are reviewed in this article. It is not the intention of this review to provide an extensive coverage of the vast amount of experimental and theoretical reports, some of which have been reviewed in several recent publications [3, 5, 18-21]. Instead, besides firmly established knowledge in this rapidly expanding field, this review pays attention to several unresolved issues in thermal and TE transport in nanomaterials, so as to stimulate further investigations into these intriguing phenomena.

## PHONON CONFINEMENT, INTERFACE SCATTERING, AND THE CASIMIR LIMIT IN SEMICONDUCTOR NANOSTRUCTURES

### Lattice Thermal Conductivity of Nanowires

The thermal conductivity is dominated by the phonon or lattice contribution ($\kappa_l$) except in metals and degenerately doped, small-band gap semiconductors. This contribution can be interpreted using

$$\kappa_l = \sum_s \int v_z(s,q)^2 \tau(s,q) \frac{k_B(\hbar\omega(s,q)/k_BT)^2 e^{\hbar\omega(s,q)/k_BT}}{\left(e^{\hbar\omega(s,q)/k_BT}-1\right)^2} \frac{dq^3}{8\pi^3} , \qquad (4)$$

where for each polarization $s$ and wave vector $q$, $v_z$ is the group velocity ($v$) component along the transport direction, $\tau = l/v$ is the relaxation time, $l$ is the phonon mean free path, $\omega$ is the frequency, and $dq^3$ is an infinitesimal volume in the Brillouin zone of the reciprocal lattice. For an isotropic bulk crystal, the integration in the $q$ space can be carried out over the phonon frequency domain as

$$\kappa_l = \sum_s \frac{1}{3} k_B \int_0^{\omega_{ZB}(s)} v(s,\omega)^2 \tau(s,\omega) \frac{(\hbar\omega/k_B T)^2 e^{\hbar\omega/k_B T}}{\left(e^{\hbar\omega/k_B T} - 1\right)^2} D(s,\omega) d\omega \qquad (5)$$

where $D(s, \omega)$ is the density of states of polarization $s$, $\omega_{ZB}$ is the zone boundary frequency, and the factor of 1/3 originates from the replacement of $v_z^2$ in Eq. (4) with $v^2$ in Eq. (5). Under the Debye approximation of a constant group velocity $v_s$, $D(s, \omega) = \omega^2/2\pi^2 v_s^3$. Phenomenological models have been reported in the literature [22-24] for calculating various scattering rates and for differentiating between the normal and umklapp processes of three-phonon scattering events.

Surface scattering of phonons in a nanowire (NW) is expected to lower $l$ and $\kappa$ to values below those for bulk crystals. Casimir [25] derived that the average boundary scattering mean free path ($l_b$) of quasi-particles equals the diameter ($d$) for a circular rod, and $1.12W$ for a rod with a square cross section of a width $W$, for the case that the quasi-particles are scattered diffusely by the surface. The results for other cross section shapes have been reported in a recent theoretical work [26]. Ziman extended Casimir's theory by defining a specularity parameter ($p$) as the probability that the quasi-particle undergoes specular reflection at the surface [27], with the lower limit $p = 0$ for diffuse scattering, and the upper limit $p = 1$ for specular scattering. For random surface roughness, Ziman obtained the following expression for a plane wave at normal incidence,

$$p(\lambda) = \exp\left(-\frac{16\pi^3 \eta^2}{\lambda^2}\right), \qquad (6)$$

where $\lambda$ is the wavelength of the energy carriers, and $\eta$ is the roughness height. Ziman suggested that this expression can underestimate $p$ for incident waves at a large or glancing angle. In this framework, the boundary scattering mean free path is obtained as

$$l_b = \alpha \frac{1+p}{1-p} l_{b,c} \ ,\tag{7}$$

where $l_{b,c}$ is Casimir's result for diffuse scattering in an isotropic material with $\alpha = 1$ and $p = 0$. For anisotropic materials, $\alpha$ can be larger than and smaller than 1 for transport along the crystallographic direction of a high and low group velocity, respectively, because of a phonon focusing effect [28].

The mean free path ($l_i$) of other intrinsic scattering processes can be combined with $l_b$ via Matthiessen's rule to obtain the overall mean free path as

$$l^{-1} = l_b^{-1} + l_i^{-1} \ .\tag{8}$$

It is worth noting that Matthiessen's rule is based on the assumption that different scattering processes are independent of each other.

With the use of a phenomenological model of Morelli *et al.* [23] that is expressed in a more detailed form than Eqs. (4-5), A. L. Moore has calculated the room-temperature thermal conductivity of Si wires as a function or the wire diameter. The result is shown in Fig. 2. In addition to the use of Casimir's result of $l_b = D$ and the same frequency-dependent umklapp scattering rate as that found in bulk Si, the calculation is based on the bulk Si phonon dispersion.

The phonon dispersion can become different from that in the bulk as the NW diameter is reduced to approach the dominant phonon wavelength. This issue has been examined by a number of studies. Some of the initial works suggested that the transition from bulk to quasi-one dimensional (1D) phonon dispersion in a 20-nm-diameter Si NWs can reduce the phonon group velocity component ($V(n,\omega)$) along the NW axis [29, 30], to be about half of the phonon group velocity $v$ in the bulk [30]. In this approach [29, 30], the bulk phonon group velocity $v$ in the expressions for the bulk scattering rates and thermal conductivity was replaced by $V(n,\omega)$ for calculating the corresponding values for the NW. Besides the enhanced boundary scattering rate because of the small NW diameter, this approach also resulted in enhanced isotope and umklapp scattering rates because of the reduced velocity value used in the calculation. In a different approach [31], the $v$ term in the thermal conductivity and scattering rate expressions was also replaced with the group velocity component along the NW axis, and the quasi-1D density of states was used in place of the $D(\omega)$ term in the bulk thermal conductivity expression of the type of Eq. (5). For the case of a 15-nm-diameter Ge NW [32], this latter approach yielded a factor of two lower thermal conductivity than that calculated with the bulk phonon dispersion.

In comparison, Mingo [33] suggested that the bulk umklapp and impurity scattering expressions are valid for the NW diameter $d > 10$ nm and at temperatures $T > 40$ K, because the energy spacing between consecutive subbands, that is $\hbar\Delta\omega \sim (2\pi/d)\hbar v$, is larger than the thermal energy $k_B T$, or the $\lambda_{p,0}$ value of 2 nm or less is still considerably smaller than $d$. In this case, the total phonon group velocity instead of just the axial component in the NW should be used in the bulk scattering expressions, according to [33]. In fact, the axial component of the phonon group velocity in a NW is that along the transport direction, and is thus equivalent to the bulk $v_z$ in Eq. (4) instead of the bulk $v$ in Eq. (5). As mentioned in the paragraph following Eq. (5), the $v_z^2$ value averaged over the entire $q$ space is one third of the average $v^2$ for the case of an isotropic bulk crystal. Recognizing that the phonon group velocity component along the NW axis is essentially $v_z$ instead of $v$, Mingo has derived the thermal conductivity expression for a quasi-1D NW, i.e. Eqs. 4-6 of ref. [33], based on the actual phonon dispersion in the NW. It is worth noting that the thermal conductivity expression obtained by Mingo [33] does not contain the 1/3 factor in Eq.

(6) of ref. [31]. For Si NWs of a diffuse surface and a diameter of 20 nm and larger, the room-

temperature thermal conductivity results calculated by Mingo [33] based on the actual dispersion are close to those calculated based on the bulk phonon dispersion, and are comparable to the calculation results shown in Fig. 2.

When the NW diameter is reduced further to be the order of the dominant phonon wavelength, the scattering rates need to be calculated based on explicit integration of scattering amplitudes amongst the confined phonon dispersions. Moreover, the frequency dependence of the scattering rates can be different between 3D, 2D and 1D systems. For example, the relaxation time approximation can be used to derive that the isotope scattering rate is proportional to $\omega^4$, $\omega^3$, and $\omega^2$ for a linear acoustic branch in 3D, 2D, and 1D systems, respectively [34].

Several measurements of Si thin films [35], Si NWs grown by a vapor-liquid-solid

(VLS) process [36], $SnO_2$ nanobelts [37], and InAs NWs [38] have yielded $\kappa$ values that can be

attributed to diffuse surface scattering alone. However, the $\kappa$ of a 22 nm diameter Si NW [36]

fell somewhat below the calculation result in Fig. 2 based on diffuse surface scattering and bulk

phonon dispersion. In addition, the thermal conductivity of this NW shows unusual linear

temperature dependence from 20 K to room temperature. In comparison, the low-temperature

thermal conductivity is expected to be dominated by boundary scattering to follow the $T$ and $T^3$

dependence of specific heat, respectively, for a 1D and 3D system. Because the 22 nm diameter of the NW is still much larger than the dominant phonon wavelength in most of the temperature range of the measurement, 1D behavior is not expected for the NW. Such linear behavior was also observed in another measurement of thin Si NWs, although only the thermal conductance but not the thermal conductivity was reported in that work because of uncertainty in the cross section [39]. One theory was able to explain the linear behavior by considering specular scattering of long-wavelength phonons at the NW surface, combined with incoherent, diffuse scattering of short-wavelength phonons [40]. However, it is unclear whether the model can also match the measured thermal conductivity magnitude. When the mean free path of short-wavelength phonons are assumed to be the same as the NW diameter, specular surface scattering of long-wavelength phonons is expected to increase the thermal conductivity.

Moreover, two recent measurements of Si NWs fabricated by electrochemical etching [41] or reactive ion etching (RIE) [42] have yielded $\kappa$ values one order of magnitude lower than the calculation result shown in Fig. 2 for diffuse surface and bulk phonon dispersion. The value reported in ref. [42] for an array of ~10 nm wide etched Si NWs supported on $SiO_2$ is even lower than the thermal conductivity of amorphous Si, that is, the amorphous limit. However, in a recent work by Hippalgaonkar *et al.* [43], the measured thermal conductivity of Si NWs etched by RIE was just slightly lower than the calculated result shown in Fig. 2, and considerably higher than those reported earlier for electrochemically etched Si NWs.

These surprising results have led to theoretical investigations of whether the Casimir limit of diffuse surface scattering can be violated. Moore *et al.* [44] employed Monte Carlo (MC) phonon particle transport simulation to investigate the effect of adding periodic sawtooth roughness to a Si NW of a 22 x 22 nm square cross section. For a NW with diffuse and flat surfaces, the MC simulation yielded similar thermal conductivity as the calculation result of Fig. 2.However, the MC simulation obtained lower thermal conductance ($G_{th}$) for NWs with large-angle sawtooth roughness than the value for flat, diffuse NW of a cross section the same as the smallest cross section of the sawtooth NW. Based on an analysis of photon scattering by a single sawtooth, Moore *et al.* [44] suspected that the added periodic sawtooth roughness can yield strong phonon backscattering, which is equivalent to a negative $p$ value. Although the backscattering effect by periodic sawtooth roughness can be used to explain the room-temperature $\kappa$ value measured for the 22 nm diameter Si NW by Li *et al.* [36], the effective $\kappa$ calculated by the MC method is still considerably higher than those reported for etched Si NWs of a much larger diameter and smaller roughness to diameter ratio.

Kazan *et al.* [45] developed a model to fit the $\kappa$ values reported for both VLS Si NWs and electrochemically etched rough Si NWs. One emphasis of their model is on scattering between optical and acoustic phonons. However, their $l_b$ term is not smaller than the Casimir result for diffuse surface scattering.

Martin *et al.* [46] developed a surface roughness scattering model based on the first-order perturbative approximation, or the Born approximation. When adding this surface roughness scattering term on top of a diffuse surface scattering term using Matthiessen's rule, Martin *et al.* [46] were able to fit the reported $\kappa$ of rough Si NWs by adjusting the surface roughness to be 3-3.25 nm. By setting the surface roughness to be close to atomic corrugation of 0.1-0.3 nm, the model can also fit the reported $\kappa$ of VLS Si NWs.

However, a recent report by Carrete *et al.* [47] suggested that the Born approximation is invalid when the surface roughness is comparable to the wavelength of phonons. Based on a Green's function calculation, these authors [47] reported that the Born approximation can overestimate the surface scattering by one order of magnitude, and that the Casimir formula yields a good approximation to the phonon mean free path. Carrete *et al.* [47] suggested that Casimir's limit in the mean free path can be overcome using large defects that penetrate deep in the NWs, but it is difficult to achieve a reduction in $\kappa$ of more than one order of magnitude below that calculated based on diffuse surface scattering and the actual phonon dispersion without compromising other properties including the structure stability. Based on the finding, Carrete *et al.* [47] emphasized that a plausible explanation for the experimental results reported for rough Si NWs remain an open question for theory.

In addition, Sadhu and Sinha [48] suggested that surface scattering has been counted twice in Martin *et al.* [46]'s calculation that included the frequency-independent diffuse surface scattering along with the frequency-dependent scattering. Sadhu and Sinha [48] proposed that coherent effects, specifically, interference of scattered waves by multiple correlated points at the NW surface of a Gaussian-type roughness distribution, can result in mean free paths for high-frequency phonons well below the Casimir limit. Moreover, Sadhu and Sinha [48] emphasized that the surface roughness correlation length needs to be less than the NW diameter in order for the surface scattering mean free path to go below the Casimir limit. Based on this theory, these authors [48] attributed the much lower thermal conductivity observed in the electrochemically etched NWs [41] to a larger ratio between the roughness amplitude and the correlation length than that of the RIE etched NWs of Hippalgaonkar *et al.* [43]. Sadhu and Sinha [48] suggested that the room-temperature umklapp scattering mean free paths of the high-frequency phonons are much longer than the attenuation length due to surface scattering in ~50-nm-diameter electrochemically etched Si NWs. In addition, it was assumed in their work [48] that the thin oxide layer at the NW surface does not play a dominant role in surface scattering, or phonon transmission into and out of the amorphous oxide shell does not greatly diminish wave interference, so that the coherent scattering effect can be observed. It was suggested that the Bourret approximation used in their calculation [48] imposes less restrictive condition on the relative magnitudes of the phonon wavelength and the surface roughness compared to the Born approximation. Nevertheless, it was pointed out in that work [48] that there is no guarantee that the Bourret solution will converge to the actual solution even if the approximation is valid, and that further studies are needed to examine scattering by non-Gaussian roughness statistics in actual NWs in the experiments.

Most recently, He and Galli [49] employed equilibrium molecular dynamics (MD) simulation to calculate the thermal conductivity of 15-nm-diameter Si NWs of different surface terminations, surface roughness, and core defects. High-temperature annealing and subsequent cooling was used in the MD simulation to reconstruct the NW surface so that the resulting surface can better match the experimental condition. For a 15-nm-diameter Si NW without surface amorphous layer or disorder, the calculated thermal conductivity was only a factor of three smaller than the bulk value. This reduction was attributed to the suppression in both the phonon lifetime and group velocity. Adding an amorphous Si or amorphous surface layer of 0.5 and 1 nm thickness leads to an additional reduction of the thermal conductivity by a factor of 3 and 4, respectively, so that the MD thermal conductivity is about one order of magnitude lower than the bulk value. These results reveal the important effect of the thin amorphous layer on phonon scattering by the surface, contracting the assumption made by Sadhu and Sinha [48]. Adding ripples of 0.5 and 1.5 nm depth to the NW with a 0.5 nm amorphous Si layer results in a reduction in the MD thermal conductivity by another factor of 1.5 [49], so that the MD thermal conductivity value would be similar to the calculation result shown in Fig. 2. In comparison, adding 1.5 nm ripples to a NW with a 1-nm-thick $SiO_2$ layer leads to a thermal conductivity that is a factor of 30 lower than the bulk value. This value would be somewhat lower than the calculation result shown in Fig. 2, and fall between the value of the value obtained by Li *et al.* [36] and that by Boukai *et al.* [42]. However, only in the NWs with grain boundaries in the NW core, the MD thermal conductivity value can be a factor of 100 lower than the bulk value and approach the data of Boukai *et al.* [42] for the ~10-nm-diameter Si NW. Based on these calculation results, He and Galli [49] proposed that core defects or large contact thermal resistance might have been present in those Si NWs where the measured thermal conductivity value was unusually low. These results suggest the need for additional experimental and theoretical studies for clarifying the relative contributions from the core defects and surface scattering.

### Nanomeshes and Phononic Crystals

Introduction of artificial spatial periodicity in the elastic properties of a material can in principle result in the appearance of bandgaps in the phonon frequency spectrum, and flattened folded optical branches of a low group velocity in a reduced Brillouin zone [50]. These effects on the phonon dispersion in the long-wavelength, low-frequency range may be observed directly in experiments. For example, a recent work reported Brillouin light scattering (BLS) measurements of low-frequency phonon dispersion even in a superlattice structures consisting of alternating porous Si layers of two different porosity values and with ~ 500 nm superlattice period [51]. Low-lying flat optical phonon branches of ~40 GHz frequency were observed by the BLS measurements. Moreover, thermal conductivity suppression measured in epitaxial superlattice structures was attributed to the reduced phonon group velocity of the folded phonon branches and the formation of phononic bandgaps in several works including ref. [52].

Recently, very low in-plane thermal conductivity values approaching the amorphous limit have been reported for periodic thin nanomesh Si membranes [53, 54]. The $\kappa$ values are even lower than those measured for etched NW arrays with a similar critical dimension. In one work [54], the thermal conductivity reduction was attributed to phonon-surface scattering at the 23 nm wide necks in the periodic holey Si membrane structure of 55 nm pitch. The measured thermal conductivity is lowered than the result of a MC phonon transport calculation [55] that considers diffuse surface scattering at the pore surface. However, the geometry considered in the MC simulation is different from that in the measurement. Hence, further MC calculations of the actual experimental structures would be helpful to verify whether the experimental values are indeed well below that caused by diffuse surface scattering alone. In another work [53], the low thermal conductivity is attributed to phononic bandgaps and reduced group velocity in the periodic nanomesh phononic crystal structure, which has a neck width of about 18 or 23 nm and a period of 34 nm. In a more recent work [56], the cross-plane thermal conductivity of 500-nm-thick Si membranes with patterned periodic pores of 300 nm pore diameter and 500 nm period was measured with the use of a transient thermal reflectance method. The room-temperature value is $5.84 \pm 1.3$ W m$^{-1}$ K$^{-1}$, about one order of magnitude lower than the result from a MC calculation [57] that assumes diffuse phonon-surface scattering. The measurement result was attributed to modification of the phonon dispersion and group velocity by the periodic phononic crystal structure.

However, the phonon wavelength spectrum is very broad. Although the spectral phonon thermal conductivity peaks at a somewhat longer wavelength than $\lambda_{p,0}$ given in Eq. (3), this wavelength is still much smaller than the period or pore size in the reported Si nanomesh structures, and may be comparable to the surface roughness of the pores. Diffuse scattering by random interface roughness can potentially destroy the phase coherence required for phononic bandgap formation. In fact, even for thin film epitaxial superlattice structures, atomic mixing at the interface is believed to result in partial diffuse scattering of short-wavelength phonons according to some theories [58]. This group of theories [58] treats phonons as incoherent particles, and considers interface scattering as the classical size effect. These models can explain the experimental data reported for superlattices of periods thicker than ~5 monatomic layers. In addition, the ~200 nm wide spacing between pores in ref. [56] can be comparable to or shorter than the room-temperature umklapp phonon scattering mean free path. Thus, umklapp scattering can destroy phase coherence and phononic bandgap formation in the high frequency range important for thermal transport at room temperature. Hence, there exist opportunities to carry out further theoretical and experimental works to investigate whether phononic bandgap formation or another effect has played a dominant role in the very low thermal conductivity observed in these Si nanomesh structures.

### Progresses and Challenges in Thermal Measurements of Individual NWs

Most of the aforementioned measurements of individual NWs and nanomeshes were made using suspended resistance thermometer (RT) devices similar to or modified from the method reported earlier by Shi [59], Kim *et al.* [60], and Shi *et al.* [61]. Figure 3 shows a scanning electron microscopy (SEM) image of one of such devices and the thermal resistance circuit. Moore and Shi [62] recently provided a numerical heat transfer analysis of the systematic measurement errors of this method for suspended and supported NWs at different temperatures. Besides the systematic errors that can be large for supported NW samples, a major measurement uncertainty originates from the contact thermal resistance ($R_c$) between the NW sample and the two micro-thermometers. The $R_c$ can be large. For example, Zhou *et al.* [63, 64] found that individual InAs NWs suspended between two micro-thermometers tended to crack near one membrane, possibly due to vibration. Such a crack was visible only in high resolution SEM images. With the presence of such a crack, one InAs NW essentially made a point contact to one of the two suspended thermometers, resulting in very large $R_c$.

In an effort to determine the $R_c$, Mavrokefalos *et al.* [65] developed a four-probe TE measurement procedure to obtain the intrinsic $\kappa$, $S$, and $\sigma$ of the same individual NW suspended between the two micro-thermometers. In this procedure, the electrical contact between the four Pt electrodes and the NW sample are used as thermocouple junctions to determine the temperature drops at the two contacts, as illustrated in Fig. 3. This method has been successfully applied for measuring $Bi_2Te_3$ NWs [66], where clean electrical contact between the NW and the underlying Pt electrodes can be made by annealing of the sample in a forming gas. However, for some NWs such as Si NWs with a stable native oxide on the surface, making clean electrical contact between the NW and the Pt electrodes on the suspended device has remained difficult, preventing the use of the four-probe TE measurement procedure.

Recently, Hippalgaonkar *et al.* [43] have patterned Si NWs together with the micro-thermometers without an interface between the Si NWs and the Si membrane support under the Pt resistance thermometers. These authors [43] measured different patterned Si NW samples of different lengths, and determined that the junction resistance between the Si NW and the Si membrane from the measured thermal resistance versus length curve. The obtained $\kappa$ of the etched Si NWs was slightly lower than the Casimir limit shown in Fig. 2, and considerably higher than those reported earlier for electrochemically etched Si NWs, as discussed above.

Besides the uncertainty due to the $R_c$, the cross section and $\kappa$ of rough NWs are non-uniform along the NW and are not well defined. The narrower sections of the NWs make a larger contribution to the measured thermal resistance than the wider sections. If the NW diameter $d$ is taken to be the peak-to-peak diameter, it can overestimate the cross section that scales with $d^2$ and underestimate $\kappa$. In addition, the Si NWs found with unusually low $\kappa$ are all degenerately doped. Hence, it will be of value to characterize the doping and defect concentrations of the measured NWs in future works and account for these factors in theoretical analysis. As suggested in a recent review [18], due to the sensitivity of single NW thermal conductivity measurements, additional independent characterization will be extremely helpful.

# PHONON TRANSPORT IN ISOLATED LOW DIMENSIONAL SYSTEMS

## Theoretical Analysis of the Thickness Dependence of the Thermal Conductivity in Low-Dimensional Systems

As the thickness is reduced below the dominant phonon wavelength, such as in singe-layer graphene (SLG) and single-walled CNTs (SWCNTs), the transition from 3D to 2D or 1D transport can reverse the thickness dependence of $\kappa$. In 2D or 1D systems, the allowable wave vector states lie on well separated planes or lines in the reciprocal space, or only one plane for the case of SLG and one line for a 1D monatomic lattice chain. Such quantization strongly restricts the phase space that can satisfy the energy and momentum conservation requirements for phonon-phonon scattering. In addition, the atomically smooth surfaces of suspended flat SLG or straight SWCNTs cannot scatter phonons diffusely because of the absence of states in the SLG or SWCNT with a velocity component perpendicular to the surface. Moreover, inter-layer scattering that reduces the basal plane $\kappa$ of graphite is eliminated in the SLG and SWCNT. Hence, it has been suggested [67, 68] that these atomically thin carbon allotropes with strong sp$^2$ bonding can possess ultrahigh $\kappa$ exceeding the previous record values found in diamond and graphite.

Because of the reflection symmetry in 2D suspended SLG, Lindsay *et al.* [69] has suggested an additional selection rule for phonon-phonon scattering processes involving the out-of-plane transverse acoustic phonons, i.e. ZA or flexural phonons. This selection rule requires that such processes need to involve an even number of ZA phonons, resulting in a very restricted phase space for umklapp scattering of the ZA phonons.

The ZA branch shows a quadratic dependence of $\omega$ on $q$, with a lower group velocity than the linear longitudinal acoustic (LA) and in-plane transverse acoustic (TA) branch. It is worth noting that the phonon branch with the lower group velocity can make a larger contribution to the specific heat and $\kappa$ at low temperatures because of higher occupation of these phonon modes. For graphene, the thermal energy $k_B T$ at room temperature is still much lower than those of LA and ZA phonons of energy higher than the zone-boundary value of the ZA branch. When the phonon mean free path is constant and the same for all phonon modes, Equation (4) can be used to show that the ZA contribution to $\kappa$ dominates the LA and TA contributions at room temperature and below in suspended flat SLG [70, 71]. The additional restrictive selection rule for ZA phonons serve to reduce ZA phonon scattering and increases its contribution to $\kappa$ in flat suspended SLG [69, 70]. This theory conflicts with another calculation based on the relaxation time approximation (RTA) [68], which does not calculate the scattering phase space and yields a large umklapp scattering rate for ZA phonons.

The calculation of Lindsay *et al.* [72] shows that the $\kappa$ of SWCNTs approaches that of flat 2D graphene at the large diameter limit. When the SWCNT diameter is reduced, the increased curvature breaks the reflection symmetry, leading to enhanced anharmonic scattering of ZA phonons and decreased $\kappa$. As the diameter is reduced further, the energy separation between adjacent 1D phonon energy subbands is increased, and optical phonons are pushed to higher frequencies. This feature lowers the scattering rate for low-$\omega$ acoustic phonons and increases their contribution to $\kappa$. These two competing effects result in a minimum in the $\kappa - d$ curve, which occurs at $d_{min} \approx 1.5$ nm for a given CNT length $L = 3$ µm based on the calculation of Lindsay *et al.* [72].

**Theoretical Analysis of the Length Dependence of the Thermal Conductivity in Low-Dimensional Systems**

Several theoretical studies have predicted that the $\kappa$ value increases and diverges with increasing lateral size of 2D and 1D solids [73-75]. Under the Debye approximation, the phonon density of states ($D(\omega)$) is proportional to $\omega^2$, $\omega$, and $\omega^0$ for the 3D, 2D, and 1D cases. Hence, as the dimension is reduced, the $D(\omega)$ for low-$\omega$ phonons is increased compared to that of high-$\omega$ phonons. Low-$\omega$ phonons with enhanced $D(\omega)$ in 1D and 2D are not subject to anharmonic scattering and are essentially ballistic carriers of heat. During a thermal measurement, when the two ends of the 1D or 2D system are in contact with two thermal reservoirs at different temperatures, the finite interface transmission coefficient ($\chi \leq 1$) at the contact limits the contribution of the low-$\omega$ phonons to a finite value of $\chi G_0$, where $G_0$ is the ballistic thermal conductance [76] of these low-frequency phonons and is independent of the sample length ($L$). Hence, as $L$ increases, the measured in-plane or axial $\kappa$ is expected to increase. However, Mingo and Broido have suggested that both higher-order umklapp phonon scattering and defect scattering can be effective to prevent the divergence [77].

It is worth noting that the long umklapp scattering mean free path of long-wavelength phonons in clean, suspended graphene and CNTs creates a challenge for classical MD calculation of thermal transport in these low-D systems, because currently MD calculation can only handle short graphene ribbons or CNTs with a length smaller than the long phonon mean free path. As pointed by Mingo and Broido [78], the thermal conductance values obtained by some non-equilibrium MD calculations of thermal transport in short, suspended CNTs violate the ballistic thermal conductance limit. A common practice to circumvent this problem is to extrapolate the MD-calculated thermal conductivity versus length data to the long length limit. However, the accuracy of this approach remains to be investigated. More importantly, classical MD calculations do not obey the Bose-Einstein statistics of phonon quasi-particles. This and other issues are expected to lead to future research opportunities in MD simulations.

### Thermal Conductivity Measurements of Suspended Carbon Nanotubes

For verifying the ultrahigh $\kappa$ and intriguing transport phenomena suggested for CNTs, different methods have been reported for thermal transport measurements of individual CNTs. Kim *et al.* [60] fabricated an earlier version of the suspended micro-RT device shown in Fig. 3 for thermal measurements of multi-walled (MW) CNTs, and reported a room temperature $\kappa$ value of about 3000 W m$^{-1}$ K$^{-1}$ for a 14-nm-diameter MWCNT, as shown in Fig 4. In this device [60] as well as a T-junction sensor [79], the thermal conductance $G_{th}$ of the CNT sample assembled in the device was obtained from a steady state measurement. The contact thermal resistance problem can be a much more serious problem for these direct G measurements of high-$\kappa$ CNTs than for low-$\kappa$ NWs. Prasher [80] has shown that the lower effective $\kappa$ measured by Kim *et al.* [60] at low temperatures in the MWCNT than in graphite was caused by a large $R_c$.

In a different approach, Lu *et at.* [81] demonstrated a 3ω method based on modulated electrical heating of the sample for measuring the thermal conductivity and the specific heat of an individual Pt wire and a MWCNT bundle. This method was used by Choi *et al.* [82] for a 1.4 μm-long MWCNT. However, the 3ω and other self electrical heating methods cannot be applied readily to SWCNTs, because selective coupling between electrons and optical phonons can result in non-equilibrium phonon populations [83]. To account for non-equilibrium transport in a current-carrying SWCNT, Pop *et al.* [84] developed a coupled electron-phonon transport model to fit the current-voltage characteristics of a SWCNT by adjusting the coupling constant between optical and acoustic phonons and contact thermal and electrical resistances. This approach was used to extract that the $G_{th}$ of a 2.6 μm-long SWCNT is inversely proportional to $T$ at above 300 K [84].

Raman spectroscopy has been explored as a non-contact technique for probing thermal transport in carbon nanomaterials. The anti-Stokes to Stokes intensity ratio of a Raman peak directly yields the temperature of the Raman-active optical phonons in the laser spot [85]. This method can be used when the temperature of the optical phonons is as high as about 600 K, below which the anti-Stokes peak is too small to be measured accurately. Alternatively, the shift in the Raman peak position can be used to probe the equivalent local temperature of phonons that are involved in anharmonic scattering of the Raman-active optical phonons [86, 87]. However, the Raman peak position also depends on strain and impurity doping [86, 88], which needs to remain constant during the thermal measurement in order to attribute the observed Raman peak shift to the temperature change alone. In addition, long-wavelength acoustic phonons can have a very long mean free path in clean, suspended CNTs and graphene. If these phonons do not interact with the photons or electrons effectively and are not involved in the anharmonic scattering processes of Raman-active optical phonons, they would be at a lower local temperature than the optical phonon temperature in optically heated or electrically biased CNTs and graphene. Indeed, non-equilibrium between different phonon populations has been observed by Raman spectroscopy in electrically biased CNTs [89-91].

There have been efforts of measuring the thermal conductance of CNTs with the use of micro-Raman spectroscopy. Hsu *et al.* [92] used the Raman peak shift to probe the local temperature and the ratio between the contact thermal resistance and the intrinsic thermal resistance in optically heated individual suspended CNTs and bundles, where phonon transport was diffusive because of defect scattering. Deshpande *et al.* [93] employed this approach to profile temperature distribution along electrically biased, high-quality suspended SWCNTs. For a 2-μm-long SWCNT, the Raman measurement revealed non-equilibrium between different phonon populations, and the Raman measurement result was used to determine the thermal conductivity contribution from the hot optical phonons with a non-negligible group velocity in the CNT. For a 5 μm-long SWCNT, Deshpande *et al.* [93] reported that different phonon populations were approximately at local equilibrium and the thermal conductance was proportional to $T$ at above 300 K. In another work, Li *et al.* [94] employed Raman spectroscopy to measure the $T$ and $G$ of current-carrying SW- and MW- CNTs where local thermal equilibrium among different phonon branches was established in the >30 μm-long CNTs. In these Raman measurements of thermal transport in CNTs [92, 93], $R_c$ could be determined from the obtained $T$ profile. However, the $G_{th}$–$T$ dependence has not been obtained partly because the temperature resolution of the Raman measurements was 50 – 100 K and the temperature rise used in the experiments usually exceeded 100 K [93, 95]. In addition, it is important to conduct the Raman measurements in vacuum so as to eliminate heat loss from electrically or optically heated CNTs to surrounding gas molecules [96]. Moreover, the radiation loss needs to be properly accounted for in Raman measurements of long suspended CNTs.

These measurements of individual SWCNTs and MWCNTs have yielded $\kappa$ values that vary by one order of magnitude, as shown in Fig. 4. The variation can be attributed to different sample qualities as well as the uncertainties in $R_c$ and the cross section. The cross section of the CNT sample was not characterized in detail in many of these measurements. The 14-nm diameter of the MWCNT sample reported by Kim *et al.* [60] was determined from a SEM measurement, which can consist of an uncertainty of several nanometers and cannot determine the number of walls of the sample. MWCNTs and double-walled CNTs (DWCNTs) appear to be much brighter than SWCNTs in SEM images because of a larger scattering cross section [97]. Based on this feature, Yu *et al.* [98] thought that their CNT sample grown by chemical vapor deposition (CVD) was a SWCNT, and estimated that the diameter was in a large range between 1 and 3 nm. The CNT diameter of several other measurements was determined with the use of atomic force microscopy (AFM) measurement, which cannot determine the number of walls.

In two experiments, Fujii *et al.* [79] and Chang *et al.* [99] employed transmission electron microscopy (TEM) to characterize the cross section of individual MWCNTs assembled on the suspended micro- RT device or T-junction sensor for thermal measurement. In these works, the effective $\kappa$ of the MWCNTs grown by arc discharge was found to increase with decreasing diameter, as shown in Fig. 5. This trend was attributed to reduced inter-layer scattering in MWCNTs with decreasing numbers of walls. However, it is unclear whether $R_c$ could have played a role on the measurement results. In particular, $R_c$ is expected to make a larger contribution to the measured thermal resistance for a MWCNT with a larger diameter and smaller intrinsic thermal resistance. In a more recent combined TEM and $G$ measurement of CNTs grown by CVD between two suspended RTs, Pettes and Shi [97] determined the $R_c$ by comparing the measured $G$ before and after Pt-C was deposited at the contact areas to the MWCNT samples. The obtained intrinsic $\kappa$ values of the CVD MWCNT samples were also found to decrease with increasing diameter. However, the diameter dependence observed by Pettes and Shi [97] correlates well with the TEM observation of increasing defect density with increasing diameter of the CVD MWCNT samples.

Despite the uncertainty in $R_c$, the $\kappa$ values reported for the several MWCNT and SWCNT samples characterized by TEM are expected to be more accurate than those without. These combined TEM and $G$ measurements yielded room-temperature $\kappa$ in the range of 300-2900 W m$^{-1}$ K$^{-1}$ for MWCNTs grown by arc discharge [79, 99], of 45-310 W m$^{-1}$ K$^{-1}$ for CVD MWCNTs [97], and about 600 W m$^{-1}$ K$^{-1}$ for CVD SWCNTs and DWCNTs [97]. Except for the CVD MWCNTs where the $R_c$ was determined [97], these values are the lower bound of the intrinsic $\kappa$ because of the presence of $R_c$.

In addition, there have been experimental efforts for verifying theoretical prediction of length-dependent $\kappa$ in CNTs. Based on micro-RT measurements, Change et al. [100] reported length-dependent $\kappa$ in defective MWCNTs, which had a phonon mean free path shorter than 100 nm. However, Pettes and Shi [97, 101] were able to fit their data with the use of a length-independent $\kappa$ and $R_c$ given by a fin resistance model. In addition, using the self-heating 3ω method, Wang et al. [102] reported length-dependent $\kappa$ of SWCNTs supported on SiO$_2$. However, the scanning thermal microscopy (SThM) measurement reported by Shi et al. [103] has revealed that most of the electrical heating in a supported SWCNT is dissipated through the SiO$_2$ substrate instead of along the SWCNT. Consequently, the temperature rise in the current-carrying SWCNT is much more sensitive to the unknown CNT-SiO$_2$ interface thermal resistance ($R_i$) than the $\kappa$ of the SWCNT. Hence, the $\kappa$ values reported by Wang et al. [102] could have considerable uncertainty due to the unknown $R_i$ as well as the uncertainty regarding whether local equilibrium was established in the electrically biased SWCNTs [104].

Hence, further development of the experimental capability for thermal measurements of individual CNTs is required to verify the theoretical predictions of length- or diameter-dependence of the CNT thermal conductivity. In particular, while even the chiral index of one measured CVD SWCNT has been obtained from selected area electron diffraction analysis [97], measuring $R_c$ of individual CNTs remains an important area for future study.

### Thermal Conductivity Measurements of Suspended Graphene

For measuring thermal transport in graphene, Balandin *et al.* [105] used a Raman laser beam to heat a rectangular SLG flake suspended across a trench. The temperature at the laser spot of the graphene was determined from the Raman G peak shift. In the initial measurement [105], the thermal conductivity was assumed to be the same for the suspended region and the supported region of the same graphene monolayer. The thermal coupling between the graphene monolayer and the underlying $SiO_2$ support at the two ends of the trench was ignored. The thicker graphite regions at the two ends of the graphene monolayer were assumed to be perfect heat sinks. The optical absorbance of the graphene monolayer was determined by comparing the Raman peak intensity value measured on the SLG and that on graphite. This method obtained an optical absorbance of about 6.5% per pass of the laser beam through the SLG. Based on the measured graphene temperature and optical absorbance, the near-room temperature $\kappa$ of suspended SLG was obtained to be $(4.84 \pm 0.44 \text{ to } 5.30 \pm 0.48) \times 10^3 \text{ Wm}^{-1}\text{K}^{-1}$ [105-107], exceeding those of diamond and graphite.

In comparison, Nair *et al.* [108] used direct transmission measurement to determine the optical absorbance of SLG to be 2.3%. In the experiment, the reflectance was ignored because it is expected to be much smaller than the absorbance based on a theory. Using this 2.3% value, Faugeras *et al.* [109] measured a $\kappa$ value of about 600 $\text{Wm}^{-1}\text{K}^{-1}$ for a SLG sample suspended over a 44-μm-diameter hole when the Raman laser beam raised the SLG temperature at the laser spot to 660 K, which was determined from the anti-Stokes to Stokes ratio. Their circular Corbino membrane geometry matches the laser beam intensity profile and heat flow direction better than the rectangular flake geometry. Because of a larger contact area with the support, in addition, the contact thermal resistance is expected to be less an issue for the circular membrane geometry than for the rectangular flake geometry. Faugeras *et al.* [109] suggested that the discrepancy between their $\kappa$ values and those reported by Balandin *et al.* [105] was caused by the different assumptions in the optical absorbance value.

In two other Raman measurements by Cai *et al.* [110] and Chen *et al.* [111], the optical absorbance of SLG samples grown by CVD and suspended over holes of 2.9-9.7 diameters was obtained to be $3.3 \pm 1.1\%$ based on the measured transmission. Based on this optical absorbance, Cai *et al.* [110] reported that the measured $\kappa$ of the suspended graphene decreases from $(2.5 \pm 1.1) \times 10^3 \, \mathrm{Wm^{-1}K^{-1}}$ to $(1.4 \pm 0.5) \times 10^3 \, \mathrm{Wm^{-1}K^{-1}}$ when the temperature of the optically heated SLG determined from the Raman G peak shift increased from about 350 K to about 500 K. The results of Chen *et al.* [111] are similar and are shown in Fig. 6 together with the basal plane thermal conductivity recommended in ref. [112] for pyrolytic graphite (PG). In addition, with the optical absorbance assumed to be 2.3% for their SLG samples exfoliated from natural graphite directly onto holes of 2.6-6.6 μm in diameter, a recent Raman measurement by Lee *et al.* [113] yielded $\kappa$ values similar to the reported values for PG.

Hence, an unresolved issue in these Raman measurements of suspended SLG is the different optical absorbance values used in different works. The recent result by Mak *et al.* [114] indicates that the optical absorbance can be larger than 2.3% for a certain wavelength range. If the relationship between the optical conductivity and the optical absorbance still obeys the theoretical prediction, the optical absorbance extracted from their result would be close to 3.3% for the 480-515 nm wavelength range of the Raman laser. In addition, polymer or tape residues on the suspended graphene can increase both the absorbance and reflectance of the sample.

However, it is unlikely that tape residues can considerably increase the Raman G band intensity used in the earlier work [105] to determine the optical absorbance. Moreover, the results obtained based on the Raman peak shift can be complicated by the effect of strain and impurity on the Raman peak position, as discussed in the preceding section. In addition, if the phonons probed by the Raman laser are thermalized locally within a scale comparable to the laser spot, the temperature distribution in the graphene can be solved from the heat diffusion equation with a Gaussian heat source term [109, 110]. The solution relates the $\kappa$ contribution from these phonons with the measured thermal resistance ($R_m$), defined as the ratio between the Raman-measured temperature rise and the optical absorbance of the graphene. If the mean free paths of these phonons are much larger than the laser beam size so that they are not thermalized locally within the laser spot and thus do not contribute fully to thermal transport , there would be an additional ballistic resistance component ($R_b$) in the measured $R_m$ [110].

Hence, the reported graphene thermal conductivity obtained by the Raman technique contains rather large uncertainty caused by the unresolved issue in the optical absorbance, the effects of strain and impurity doping, low temperature sensitivity of more than 20-50 K of the Raman thermometry technique, and the ballistic thermal resistance. Moreover, these Raman measurements have not been employed to probe the graphene thermal conductivity at low temperatures. It is worth noting that the temperature in the Raman data of Fig. 6 is the hot side temperature in the graphene probed by the Raman data, whereas the cold side temperature was kept at near room temperature. This issue makes it nontrivial to compare the obtained graphene thermal conductivity from the Raman technique with those reported for graphite, where a small temperature drop across the sample was used during the thermal measurement and the temperature reported is usually the average temperature of the sample. These issues suggest that there are abundant opportunities to further develop the Raman and other experimental techniques for thermal conductivity measurement of clean suspended graphene samples.

# INTERACTION BETWEEN PHONONS IN LOW DIMENSIONAL SYSTEMS AND THE ENVIRONMENT

## Effect of Interface Interaction on Thermal Conductivity of Nanotubes and Graphene

Low-dimensional materials such as CNTs and graphene are often supported on a substrate for device applications or embedded in a matrix to form functional nanocomposites. Electron transport in low-D materials such as graphene and CNTs is known to be highly sensitive to the environment. For example, the electrical conductance and Seebeck coefficient of CNTs were found to be very sensitive to adsorbed gas molecules [115]. The electron mobility in graphene supported on an oxidized Si wafer is considerably lower than those for clean suspended graphene [116] and graphene supported on a clean hexagonal boron nitride (h-BN) layer [117].

Several theoretical studies [118, 119] have suggested that the $\kappa$ of CNTs embedded in a medium can be rather different from those of free-standing and clean CNTs because of scattering of phonons and even modificiation of the phonon dispersion in the CNTs. Experimental verification of these theoretical predicitons for CNTs has remained a challenge. However, the Raman measurement of Cai *et al.* [110] has obtained that the near room-temperature $\kappa$ of the gold-supported region of their CVD SLG sample was about $(370 + 650/-320)$ W m$^{-1}$ K$^{-1}$, considerably lower than the $(2500 + 1100/-1050)$ W m$^{-1}$ K$^{-1}$ value measured in the suspended region of the same sample. Moreover, Seol *et al.* [70] fabricated a suspended resistance thermometer micro-device for $\kappa$ measurements of SLG supported on amorphous SiO$_2$, and found a room-temperature $\kappa$ value of about 600 W m$^{-1}$ K$^{-1}$, which is considerably lower than those reported for PG and suspended SLG, as shown in Fig. 6. Raman spectra measured on the supported SLG samples show no presence of the D peak caused by defects. The mobility extracted from the measured $S$ and $\sigma$ is comparable to the highest value reported for SLG supported on SiO$_2$. Because these results suggest high-quality SLG samples used in their thermal measurement, Seol *et al.* [70] attributed the lower $\kappa$ of the supported SLG to SLG-SiO$_2$ interaction.

The interface interaction can impact phonon transport in supported graphene in different ways. The first one is to modify the phonon dispersion of the supported graphene. Although the phonon dispersion of graphene grown on Ni and Ru is altered considerably by the strong bonding to the metal substrate, the acoustic phonon dispersion is similar for suspended graphene, graphene physisorbed on Pt, and for graphene grown on Ni after Cu, Ag, or Au is intercalated under the graphene [120-125]. In addition, inelastic neutron scattering of ordered pyrolytic graphite [126] and electron energy loss spectroscopy of thin randomly stacked graphite [127] have shown the phonon dispersion of these materials is similar to 3D Bernal graphite [128] and 2D graphene, respectively, with the observation of the ZO′, TO′, and LO′ modes only in the ordered pyrolytic graphite sample. Thus, weakly interacting graphitic layers had been modeled in the literature with dispersions calculated for 2D graphene [129, 130]. Therefore, modification of the phonon dispersion is not expected to play a major role in the observed reduced thermal conductivity in supported graphene, compared to scattering of graphene phonons by the support.

The support can scatter phonons in graphene via several mechanisms. The interface coupling can lead to local perturbation in the diagonal force constant on the graphene atoms. In addition, the atoms in the SiO$_2$ support are not static. The interaction between vibrating atoms at two sides of the interface results in phonon scattering across the interface, although there is no net energy transfer across the interface when the temperature gradient is parallel to the interface. Perturbation theory was employed in ref. [70] to obtain the total scattering rate due to both scattering across the interface and the local perturbation of the diagonal force constant on the graphene atoms. The obtained substrate scattering rate increases rapidly with decreasing phonon frequency. This frequency dependence is similar to that for the phonon transmission coefficient across a weak van der Waals interface [131]. In addition, the interface force constant is stronger for the ZA modes than for the in-plane modes when the support is amorphous [70]. When this factor and the rather conformal morphology of supported graphene are taken into account, the substrate scattering model of ref. [70] can explain the measured $\kappa$ versus $T$ relation for the supported SLG, and suggests much stronger substrate damping of the ZA phonon lifetime than for the in-plane modes. Consequently, the ZA contribution to the thermal conductivity is smaller than those from the LA and TA branches in the supported SLG [70]. The dominant role of the substrate in damping the ZA modes in graphene instead of the in-plane modes is also observed in a recent MD simulation [132]. In a follow-up MD calculation of thermal transport in a CNT supported on $SiO_2$ [133], the low-frequency phonon lifetime was found to be suppressed most noticeably by the support. This finding is in qualitative agreement with the frequency dependence given by the substrate scattering model of ref. [70]. Moreover, artificially freezing the atoms in the $SiO_2$ support in the MD simulation was found to have negligible effect on the CNT phonon lifetime. This latter result suggests that substrate scattering of phonons in CNTs is dominated by static perturbation of the potential in the CNT. However, the calculation with frozen $SiO_2$ atoms was not included in the earlier reported MD simulation of supported graphene [132]. It remains to be investigated whether the conformal morphology and high interface thermal conductance [134] of supported graphene may increase the role of phonon scattering across the interface compared to the case of supported CNTs. In addition, interaction with the support also breaks the reflection symmetry of graphene, and thus relaxes the restrictive selection rule for three-phonon scattering involving the ZA phonons. This effect has not been accounted for by the substrate scattering model of ref. [70].

It is of considerable interest to employ the sensitive micro-resistance thermometer devices to obtain the accurate $\kappa$ versus $T$ relation of suspended graphene. Recently, some successes have been reported in the assembly of a 0.5 μm-long suspended SLG [135] and a 1 μm-long suspended 5-layer graphene [136] between two suspended micro-thermometers. The measured $\kappa$ exhibits a $T^{1.5}$ dependence at temperatures lower than 150 K for these two suspended samples, and was attributed to a dominant contribution from the ZA modes. However, the obtained room-temperature $\kappa$ of these samples is lower than 225 W m$^{-1}$ K$^{-1}$, considerably lower than the theoretical prediction for the ZA contribution.

Meanwhile, Pettes *et al.* [71] developed a method to assemble suspended bi-layer graphene (BLG) samples on suspended resistance thermometer devices. The measured $\kappa$ values of two suspended BLG samples are slightly higher than the results of Seol *et al.* [70] for SLG supported on SiO$_2$, and considerably lower than the Raman measurement results and theoretical predictions for clean, flat, suspended graphene of similar dimensions, as shown in Fig. 6. An approximately $T^{1.5}$ dependence can be observed in the $\kappa$ data of the BLG samples at temperatures below 125 K.

In 2D and the low $T$ limit, if the phonon scattering rate $\tau_p^{-1}$ is proportional to $\omega^\alpha$, Equation 4 can be used to show that the contribution to $\kappa$ is proportional to $T^{2-\alpha}$ for both the quadratic ZA branch and the nearly linear LA and TA branches in graphene. If the phonon mean free path is constant because edge or grain boundary scattering is the dominant phonon scattering process, the LA and TA contribution to $\kappa$ is proportional to $T^2$, while the ZA contribution is proportional to $T^{1.5}$ at low temperatures. Although a dominant ZA contribution and dominant phonon-edge scattering can lead to the $T^{1.5}$ dependence, Pettes *et al.* [71] calculated that the phonon mean free path needs to be much smaller than the sample lateral dimension or reported grain size of similar graphene samples in order to match their measured $\kappa$ or those reported in [135, 136]. Hence, instead of this mechanism, Pettes *et al.* [71] attribute their result to phonon-scattering by a residual polymeric layer that can be clearly observed by TEM on the suspended BLG. These authors [71] noted that the umklapp scattering with a positive $\alpha$ on the LA and TA modes together with stronger substrate scattering on the ZA modes give rise to an approximately $\kappa \propto T^{1.5}$ behavior at low temperatures in the calculation result of Lindsay *et al.* [70] for the SLG supported on SiO$_2$. Hence, the $T^{1.5}$ behavior observed in their BLG samples with polymer residues could be caused by a similarly complex scattering mechanism. Nevertheless, it is also possible that the polymer residue was distributed on the graphene surface as separated patches of lateral size larger than the dominant phonon wavelength, so as to give rise to geometric scattering with a constant mean free path. In this case, the ZA contribution can be larger than the LA and TA contribution, resulting in the $\kappa \propto T^{1.5}$ behavior at low temperatures. In either case, while the finding of Pettes *et al.* [71] is useful for the design of graphene-polymer nanocomposites for thermal management, it also points to the need of obtaining ultra-clean suspended graphene samples for fundamental studies of the intrinsic thermal properties of 2D graphene.

In addition, Jang *et al.* [137] used a different resistance thermometry method to measure the in-plane $\kappa$ of few-layer graphene (FLG) embedded in SiO$_2$. These authors [137] reported that the thermal conductivity of the encased FLG increases with the number of layers, as shown in Fig. 7. This result suggests stronger $\kappa$ suppression of the graphene-SiO$_2$ interface interaction than inter-layer interaction in FLG. In comparison, Ghosh *et al.* [138] reported that the $\kappa$ of suspended FLG decreases with increasing number of layers because of interlayer interaction, and approach the graphite limit when the number of layers increases to about ten layers.

### Interface Thermal Conductance of Nanotubes and Graphene

Besides the effects of interface interaction on phonon transport in low-dimensional materials, another property of importance for practical applications is the interface thermal conductance ($g_i$) between the low-D materials and the environment. Transient laser reflectance has been employed to determine that the $g$ at CNT-micelles interface is about $12 \times 10^6$ W m$^{-2}$ K$^{-1}$ and rather small [139]. Electrical breakdown measurements [140, 141], SThM [103], and suspended thermometer devices [97] have been used to determine this property for CNT-solid interfaces. The values obtained from different studies vary over a wide range because of variation in the surface roughness and interface adhesion energy [80, 131], which makes it difficult to determine the contact area accurately.

Interface thermal transport across graphene has been probed with the use of different methods. For graphene-SiO$_2$ interface, the measured $g_i$ value is (83-178) $\times 10^6$ W m$^{-2}$ K$^{-1}$ based on a steady state resistance thermometry measurement of encased graphene [134], and (20-110) $\times 10^6$ W m$^{-2}$ K$^{-1}$ for supported graphene exfoliated on SiO$_2$ measured by a transient layer reflectance technique [142]. A Raman measurement yielded $g_i$ of (28 + 16/-9.2) $\times 10^6$ W m$^{-2}$ K$^{-1}$ for graphene-Au interface [110]. Another transient laser reflectance measurement obtained 25 $\times 10^6$ W m$^{-2}$ K$^{-1}$ for the Au/Ti/graphene/SiO$_2$ interface [143]. Although these values are somewhat different, the variation is small compared to those reported for CNT-solid interfaces.

# INTERFACE SCATTERING OF ELECTRONS

## Effect of Surface Scattering on Charge Mobility of Nanowires

Interface scattering of electrons is a complicated process that depends on the interface roughness and charge states. Although NWs provide a model system to investigate the classical size effect of interface scattering of electrons, measuring the electron mean free path in the NW geometry has remained a challenge. Hall measurement has been established for measuring the mobility of thin films. However, this method cannot be applied readily to the NW and CNT geometry. Instead, electron mobility has been extracted from field effect measurement results of CNT transistor devices [144]. However, this method has limited use for degenerately doped NWs explored for TE applications because of the screening of the field effect by the high electron concentration in the NWs. The thermopower itself has been used to probe the electronic structure in bulk semiconductors [145] and in semiconductor junctions [146]. Based on a similar principle, Moore and co-workers [66, 147-149] have employed a two-band or one-band model to extract the Fermi level, carrier concentration and mobility from the measured $S$ and $\sigma$ of individual NWs. The as-obtained mobility for some InSb NWs [149] and $Bi_2Te_3$ NWs [66] was found to be reduced considerably compared to the bulk values. The mobility reduction is comparable to the $\kappa_l$ suppression measured on the same $Bi_2Te_3$ NWs [66]. For InSb NWs [149] the mobility was reduced more than for the lattice thermal conductivity. These results may be caused by surface charge states or surface depletion in these NWs, where the small $m^*$ results in a small concentration of charge carrier that is insufficient to screen the surface charges.

For large-$m^*$ semiconductor NWs, the charge carrier concentration is large, and the bulk electron mean free path is often very short and smaller than the typical NW diameters in the 30-100 nm range. Surface charges may also be effectively screened by the large carrier concentration in large-$m^*$ NWs. Consequently, little mobility reduction has been reported in CrSi$_2$ NWs of $m^* \approx 5m_0$ [147]. Similar reasons could have led to much less reduction of charge carrier mobility than $\kappa_l$ suppression in Si NWs [41, 42].

**Effect of Surface Scattering on Seebeck Coefficient of Nanowires**

Besides impact on charge mobility, increased interface scattering can influence the Seebeck coefficient due to effects such as that on the $r$ exponent in the energy dependence ($E$) of the charge carrier scattering relaxation time, $\tau \propto E^r$. The Seebeck coefficient consists of a diffusion component ($S_d$) and a phonon drag contribution ($S_p$) [17], i.e.

$$S = S_d + S_p \qquad (9)$$

The diffusion component for a single parabolic band in 3D systems is given as

$$S_d = \frac{1}{q}\left( k_B \frac{\left(r + \frac{5}{2}\right)F_{r+3/2}(\xi)}{\left(r + \frac{3}{2}\right)F_{r+1/2}(\xi)} - \frac{\zeta}{T}\right) \qquad (10)$$

where $q$ equals $e$ for holes and $-e$ for electrons, $e$ is the elemental charge, $\zeta$ equals ($E_f - E_c$) for electrons and ($E_v - E_f$) for holes, $E_f$, $E_c$, and $E_v$ are the Fermi level, conduction band edge, and valence band edge, respectively, and $F_n(x)$ is the Fermi-Dirac integral of order $n$ [17]. In the metallic limit of a large $\zeta$, the expression can be simplified to be a linear function of $r$ and $T$ according to:

$$S_d = \frac{\pi^2}{3}\frac{(r + \frac{3}{2})}{q\zeta}T \qquad (11)$$

For non-degenerate bulk semiconductors with $\zeta < -2k_B T$, the diffusion component of a single carrier type is reduced to a relatively weak function of the $r$ exponent as

$$S_d = -\frac{1}{q}\left(\frac{\xi}{T} - k_B\left(r + \frac{5}{2}\right)\right) \qquad (12)$$

Herring [150] considered preferential momentum transfer from electrons to phonons along the direction of the current flow in an isothermal semiconductor, and obtained an additional phonon heat flux besides the electronic heat flux that was considered in the derivation of Eq. (12). These two heat flux terms are normalized by the current density to obtain the phonon drag and electronic contributions to the Peltier coefficient ($\Pi$), namely $\Pi_p$ and $\Pi_e$. Based on the Kelvin relation, Herring [150] obtained:

$$S_p \approx \frac{e\beta v_p l_p}{q\mu T} \qquad (13)$$

where $v_p$ and $l_p$ are the group velocity and mean free path of those long-wavelength phonons with a wave vector of the order of the Fermi wave vector that scatter charge carriers, $\beta$ is a parameter that characterizes the strength of the charge carrier-phonon scattering, and $\mu$ is the mobility of the charge carriers.

As shown in Fig. 8, the $S_p$ component can be large and give rise to a large peak in the $S$-$T$ curve for temperatures between 30 and 150 K in p-type Si with doping concentration of $1\times10^{18}$ cm$^{-3}$ or lower [151], where $l_p$ is long and $\mu$ is limited by charge carrier-phonon scattering. At lower temperatures, the phonon population is small and $l_p$ is limited by boundary scattering, so that $S_p$ is small. At higher temperatures, $l_p$ and $S_p$ are reduced by increasing phonon-phonon scattering with increasing $T$. In highly doped Si, $l_p$ is reduced considerably by impurity scattering and charge carrier scattering, the latter of which also leads to momentum transfer back from phonons to charge carriers. These effects result in a small phonon drag contribution observed in p-type Si crystals with a doping concentration of $1 \times 10^{19}$ cm$^{-3}$[151], where $S$ is dominated by $S_d$. In addition, decreasing grain size is known to reduces $l_p$ and $S_p$ [150].

Hence, it is surprising that Boukai *et al.* [42] observed a pronounced $S$ peak due to phonon drag at $T$ near 200 K in a 20 nm Si NW doped with a high boron concentration of $3\times 10^{19}$ cm$^{-3}$. These authors [42] attributed this result to a dimension crossover from 3D to 1D for those long-wavelength phonons that scatter charge carriers. It was suggested that the wavelengths of these phonons are longer than about 3 nm and comparable to the NW diameter, so that there are few transverse wave vector states into which the NW boundary can scatter these phonons. In comparison, the spectral energy density peaks at a phonon wavelength much smaller than the 20 nm diameter of the NWs, based on Eq. (3). Boukai *et al.* [42] suggested that the dimension crossover of the long-wavelength phonons responsible for the phonon drag still permits the very low measured $\kappa$ values because of surface scattering of the shorter-wavelength phonons.

These findings are indeed very intriguing, and have stimulated further theoretical and experimental studies to better understand the phonon drag effect in nanostructures. A recent report [152] has suggested observable phonon drag contribution in the measured $S$ of the inversion layer of a 20-nm-thick, 420-nm-wide Si nanoribbon where the carrier density was tuned by a back gate voltage applied to the Si substrate underneath the nanoribbon. However, the measured $S$ at temperature 300 K, 200 K, and 100 K decreases with decreasing temperature, and does not show a pronounced phonon drag peak at low temperature. Based on the difference between the measured $S$ and the calculated $S_d$, the relative contribution of the phonon drag was determined to increase from 8.3% at 300 K to 46% at 100 K. These values are much smaller than those reported for the etched Si NW of a 20 nm $\times$ 20 nm cross section [42], as well as those in Si crystals with a comparable doping concentration of order $10^{18}$ cm$^{-3}$ or lower [151].

For n-type bulk Si of a doping concentration of $2.7\times 10^{19}$ cm$^{-3}$ [151] and $1.7\times 10^{19}$ cm$^{-3}$ [153], respectively, the measured $S$ data were found to decrease linearly with $T$ and extrapolates to zero at $T = 0$ K, resembling the metallic behavior given by Eq. (11). Boukai *et al.* [42] observed a similar linear metallic behavior for patterned p-type Si NWs of doping concentration higher than $7 \times 10^{19}$ cm$^{-3}$. In comparison, although a linear *S-T* behavior was reported by Hochbaum *et al.* [41] for a highly-doped 48 nm electrochemically etched p-type Si NW, the *S-T* curve extrapolates to a negative value of a rather large magnitude at $T = 0$ K, as shown in Fig. 8.

In an attempt to understand the different *S-T* trends observed in the bulk Si and Si NWs, the $S_d$ for p-type Si of different boron doping concentrations has been calculated and shown in Fig. 8. The calculation is based on a two band model that accounts for the contributions from both electrons and holes. For different doping concentrations and temperatures, the Fermi energy is determined based on charge neutrality and Fermi-Dirac statistics without approximating it as the Boltzmann distribution. The *r* value suggested by Herring [150] has been used, along with a mobility ratio of two between electrons and holes. The model reproduces the reported $S_d$ results of a single band model for non-degenerate Ge [150] and Si [151]. The calculated $S_d$–$T$ curves in Fig. 8 at $T > 200$ K show a smaller slope than the NW measurement data, and a divergence at the low $T$ limit. The divergence is caused a Fermi level settled near the middle between the band edge and the dopant energy level in the carrier freeze-out regime. In comparison, Geballe and Hull's [151] measurement result for a Si crystal of $1.8 \times 10^{18}$ cm$^{-3}$ boron doping decreases with $T$ and shows a sign change at low $T$, as shown in Fig. 8. This behavior in the carrier-feezeout regime has been attributed to impurity band conduction [151], which yields an additional contribution to $S_d$ of an opposite sign. Neither the impurity conduction band mechanism nor the phonon drag contribution has been accounted by the calculation results in Fig. 8. It remain to be investigated whether the large slope shown in the reported *S-T* curve of the highly-doped 48 nm electrochemically etched p-type Si NW is caused by these intriguing transport effects or other experimental factors. Similarly, further theoretical studies are also needed for better understanding the measured $S$–$T$ data of the polycrystalline Si samples [154] shown in Fig. 8.

In addition, although the $S$ of individual CNTs and NWs can be conveniently measured with electrodes and resistance thermometer lines patterned on the CNT or NW [155, 156], it is worth noting that the average temperature measured by the thermometer lines can be different from the temperature at the contact point between the thermometer line and the NW. It would be necessary to use a numerical heat transfer analysis to evaluate and compensate for this error [148]. Moreover, for measuring the $S$ of an array of NWs between two thermometer lines [42], the excitation current used to measure the resistance of one of the thermometer lines can leak through the NW array to the other thermometer line and return to form an unwanted current loop. This issue can cause uncertainty in the measured thermometer line temperature, especially when the electrical resistance of the NW arrays is not much larger than that of the thermometer line.

## THERMOELECTRIC TRANSPORT IN LOW DIMENSIONAL CONDUCTORS

In 1993, Hicks and Dresselhaus [157, 158] proposed that the sharply peaked electronic density of states (DOS) in 2D quantum wells and 1D quantum wires can be used to increase $S^2\sigma$ and $ZT$. These works stimulated a number of subsequent theoretical studies. In one of them, Broido and Reinecke [159] suggested that the $ZT$ enhancement in realistic quantum well thin film superlattice systems can be limited by several factors, including electron tunneling through the barrier layers that modifies the DOS and limits the power factor, increased charge carrier-phonon scattering rates and reduced charge carrier mobility, and parasitic thermal conduction in the barrier layers. Broido and Reinecke [159] further suggested that freestanding structures such as freestanding NWs may be needed to overcome these limitations in order to obtain a large $ZT$.

Several other calculations by Lin *et al.* [160] and Mingo [161] predicted that *ZT* larger than 2 at 77 K and 1.5 at 300 K, respectively, is possible in Bi and InSb NWs with diameter smaller than 10 nm. In addition, the calculation by Rabina *et al.* [162] suggested that *ZT* larger than 2 at 77 K could be achieved in $Bi_{1-x}Sb_x$NWs of diameter less than 20 nm. These materials all possess small *m\** and thus long $\lambda_e$, giving rise to quasi-1D DOS in NWs of diameter smaller than 20 nm. Most recently, Kim *et al.* [163] suggested that the power factor contribution from each conduction channel in 1D systems can be only 12% and 40% higher than the corresponding ones in 2D and 3D, respectively, so that the thickness of quantum wells and wires must be very small to achieve a higher power factor in lower dimensions after the power factor per channel is renormalized into power factor per cross section.

Many of these calculations are based on the assumption that there is no mobility reduction due to diffuse scattering of electrons. Without accounting for surface charge states, Ziman's expression can yield a larger *p* for long-$\lambda_e$ electrons in small-*m\** semiconductors than for phonons with a short $\lambda_p$. Hence, it is in principle possible to suppress the phonon mean free path more than the electron mean free path in NWs of semiconductors with a small-*m\**.

These theoretical works have motivated a number of experimental studies of thermoelectric transport in thin film and NW structures. There have been several experimental reports of *ZT* enhancement in thin film superlattice structures [164, 165] and NWs [41, 42]. However, the enhancement has been attributed mainly to suppressed $\kappa_l$ instead of enhanced $S^2\sigma$. In addition, for the case of thin film superlattices [164, 165], the *ZT* enhancement was reported for the cross-plane transport direction instead of the in-plane direction considered in earlier theoretical studies. As mentioned in the section on phononic crystals, interface scattering of phonon quasi-particles is believed to be the main cause of the reduced $\kappa_l$ in the cross-plane direction of thin film superlattices, although other intriguing wave behaviors such as phonon mini bands and phonon localization in thin film superlattices have also been discussed [166, 167].

In addition, Heremans *et al.* [168] measured large $S$ in Bi NWs, and attributed the result to semi-metal to semi-conductor transition due to the lifting of the band edges of 1D energy subbands. However power factor enhancement has not been observed in these existing experimental works on quantum wire or well structures, likely because interface charge states and interface scattering suppress $\mu$ and $\sigma$. One notable exception is the extraordinary electron mobility reported by Shim *et al.* [169] for a 120-nm-diameter single-crystal Bi NW, where a reduced carrier concentration was reported and could be caused by semi-metal to semiconductor transition. Another recent report [152] suggested a significantly enhanced power factor in the inversion layer of a 20-nm-thick Si nanoribbon where a gate voltage was applied to the Si substrate under the ribbon to tune the carrier concentration in the ribbon. In that work [152], no mobility suppression was found in a wider ribbon of the same thickness because of the lack of surface disorders and interface trapped charges compared to etched Si NWs. The power factor enhancement was attributed to confinement of the essentially two-dimensional electronic gas (2DEG) in the inversion layer, as well as an appreciable phonon drag contribution discussed above. However, because the inversion layer is one order of magnitude smaller than the thickness of the nanoribbon, the effective power factor of the entire ribbon would be still rather low. Nevertheless, this work [152] has suggested the feasibility of employing the field effect to tune the Fermi energy of electrons confined in nanometer or atomic layers for experimentally verifying the different theories on quantum enhancement of the thermoelectric power factor.

## SUMMARY

Significant progresses have been made in both theoretical and experimental studies of the quantum and classical size effects on thermal and TE properties. However, a number of fundamental questions on thermal and TE transport in nanostructures and low-D systems have remained elusive. One of the outstanding questions is whether and how much the $\kappa_l$ of NWs and nanomeshes can be reduced below that limited by the Casimir result of diffuse surface scattering. In addition, experimental results are still lacking for verifying the predicted $\kappa_l$ dependence on the length of clean suspended CNTs and graphene and on the diameter of the CNTs. For TE transport, detailed experimental and theoretical studies are needed to better understand whether and how much the power factor can be enhanced in quantum wells and wires, and whether large phonon-drag thermopower can indeed be observed in NWs and other low-D systems.

High-fidelity measurement techniques are critical for addressing these outstanding questions. It is worth noting that the measurement of the three properties entering the $ZT$ expression is more challenging for individual nanostructures and thin films than for bulk crystals, which already requires care to address a number of issues. This challenge can lead to large uncertainty and discrepancy in the measured properties. For example, recent measurements have found neither $S^2\sigma$ enhancement of the power factor [170] nor $\kappa_l$ suppression [171] in PbSeTe/PbTe nanodot thin film superlattices, for which enhanced $ZT$ was reported in an earlier report [165]. Hence, independent verification of reported measurement properties of individual nanostructures and thin films would benefit a broad research community.

For $\kappa$ measurements, techniques for eliminating the contact thermal resistance and for correcting the parasitic radiation loss at high temperatures are particularly important. Although $\sigma$ and $S$ measurements are not as complicated as for $\kappa$, sufficient cares are still necessary. As a simple example, even for four-probe electrical measurement of individual NWs, the width of the two voltage electrodes can introduce considerable uncertainty in the NW length and the obtained electrical conductivity when the width is not small compared to the NW length between the two voltage electrodes. Moreover, although electric field effect has proved to be a promising approach to tuning the carrier concentration in the nanostructure sample during thermoelectric measurements, new methods are needed to control the impurity concentration and surface charges and surface roughness so that the delicate quantum confinement effects are not obscured by these non-ideal factors.

Besides challenges in experimental study of thermal and TE transport in the nanoscale, current theoretical models can only deal with ideal systems with small number of atoms or simple band structures, and cannot account for the larger size or more complex lattice or electronic structures that have been investigated in experiments. The disconnect between current experimental and theoretical capabilities suggests abundant future research opportunities to reach down to clean low-dimensional samples in experiments and handle large and complex nanostructures in theoretical models.

**Acknowledgement**

The author acknowledges collaborations and helpful discussions with a number of colleagues for related research. A partial list includes Alex Balandin, David Broido, David Cahill, Gang Chen, Steve Cronin, Chris Dames, Mildred Dresselhaus, I-Kai Hsu, Rui Huang, Song Jing, Insun Jo, Philip Kim, Deyu Li, Xiaoguang Li, Lucas Lindsay, Arun Majumdar, Anastassios Mavrokefalos, Paul McEuen, Alan McGaughey, Natalio Mingo, Arden Moore, Eric

Pop, Michael Pettes, Ravi Prasher, Rodney Ruoff, Jaehun Seol, G. P. Srivastava, Zhen Yao, Laura Ye, Choongho Yu, and Feng Zhou. The author would like to thank Kenneth Goodson for his encouragement and editorial support during the preparation of this manuscript. The author's work in this area has been supported by the Thermal Transport Processes Program of National Science Foundation, Office of Naval Research, and Department of Energy Office of Basic Energy Science. Opinions and views expressed in this review do not necessarily reflect those of the funding agencies or the aforementioned colleagues. Although it can be expected that mistakes will be found in this review in the future, the objective of this review is to stimulate further discussions of several intriguing fundamental questions, as supposed to diminish the different viewpoints that are either included or not included in this review.

## NOMENCLATURE

| | |
|---|---|
| $E$ | energy [J] |
| $g_i$ | thermal interface conductance [W $m^{-2}$ $K^{-1}$] |
| $G_{th}$ | thermal conductance [W $K^{-1}$] |
| $G_0$ | ballistic thermal conductance [W $K^{-1}$] |
| $k_B$ | Boltzmann constant |
| $l$ | mean free path [m] |
| $m*$ | effective mass [kg] |
| $m_0$ | electron rest mass [kg] |
| $N$ | number of modes |
| $p$ | surface specularity parameter |
| $R_c$ | contact thermal resistance |

$S$      Seebeck coefficient [V K$^{-1}$]

$T$      temperature [K]

$v$      velocity [m/s]

$Z$      thermoelectric figure or merit [K$^{-1}$]

## Greek

$\kappa$      thermal conductivity [W m$^{-1}$ K$^{-1}$]

$\sigma$      electrical conductivity [$\Omega^{-1}$ m$^{-1}$]

$\lambda$      wavelength [m]

$\hbar$      reduced Planck constant

$\tau$      relaxation time [s]

$\zeta$      chemical potential measured from the band edge [J]

$\chi$      transmission coefficient

$\Pi$      Peltier coefficient [V]

## Subscripts

$b$      boundary scattering

$c$      contact or conduction band edge

$d$      diffusion

$f$      Fermi level

$i$      intrinsic scattering

$l$      lattice

$p$      phonon or phonon drag

*u*      umklapp scattering

*v*      valence band edge

**Figure captions**:

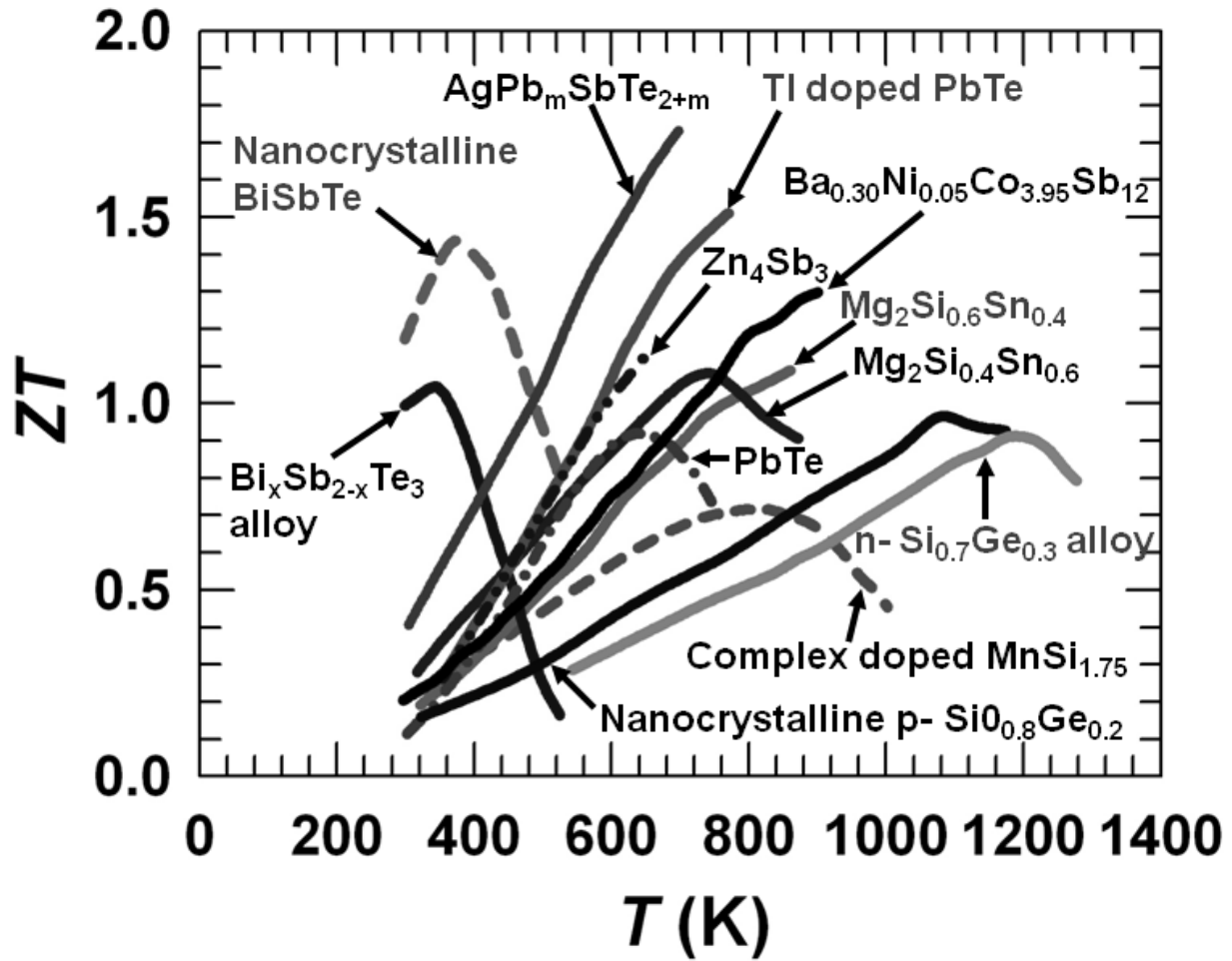

Figure 1. *ZT* as a function of temperature for bulk TE materials reported in [7-13].

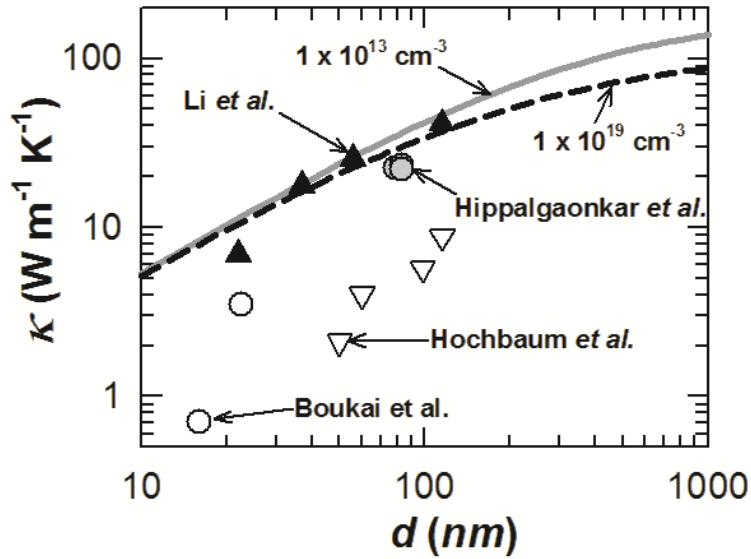

Figure 2. Experimental room-temperature thermal conductivity data of Si NWs of different equivalent diameters reported by Li *et al.* [36] for VLS NWs, Hippalgaonkar *et al.* [43] for patterned NWs, Hochbaum *et al.* [41] for electrochemically etched NWs, and Boukai *et al.* [42] for patterned NWs. The lines are the thermal conductivity of Si wires of different diameters calculated by A. L. Moore in the Casimir diffuse surface limit and with a boron doping concentration of $1 \times 10^{13}$ and $1 \times 10^{19}$ cm$^{-3}$, respectively.

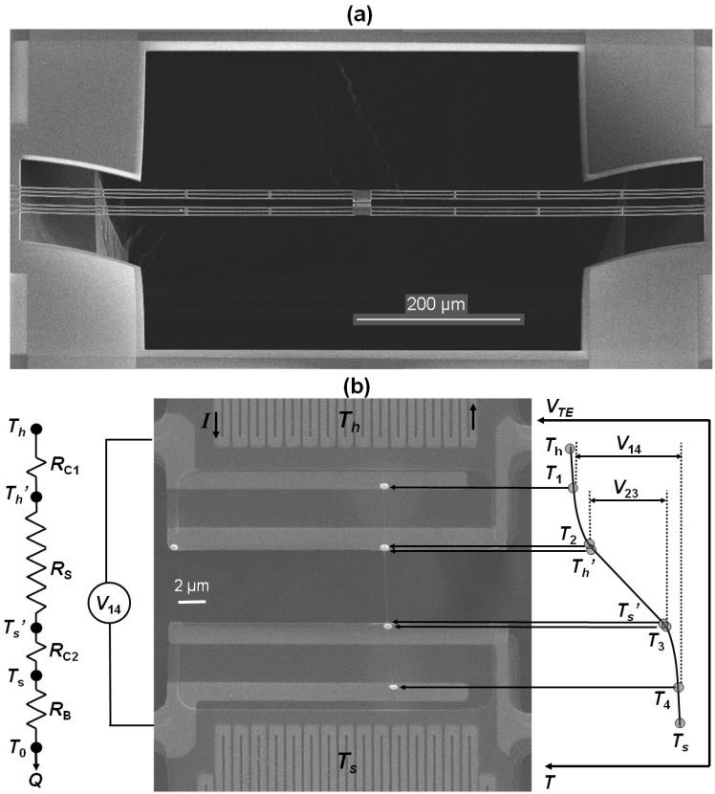

Figure 3. (a) Scanning electron micrograph (SEM) of a suspended resistance thermometer micro-device. (b) SEM of a Si NW assembled between the two central $SiN_x$ membranes of the micro-device. The thermal resistance circuit and the temperature distribution along the NW sample are shown in the schematics at the left and right sides of the SEM, respectively. $T_h$ and $T_s$ are the temperatures of the heating (upper) and sensing (lower) membranes. $T_h$' and $T_s$' are the temperatures at the two ends of the suspended segment of the film sample. $T_1$, $T_2$, $T_3$, and $T_4$ are the temperatures at the four contacts to the NW. $T_0$ is the temperature of the substrate. $R_s$ and $R_B$ are the thermal resistances of the NW sample and the six beams supporting one membrane. $R_{C1}$ and $R_{C2}$ are the contact thermal resistances between the sample and the heating and sensing membranes. $V_{14}$ and $V_{23}$ are the TE voltages ($V_{TE}$) measured between the two outer and between the two inner electrodes.

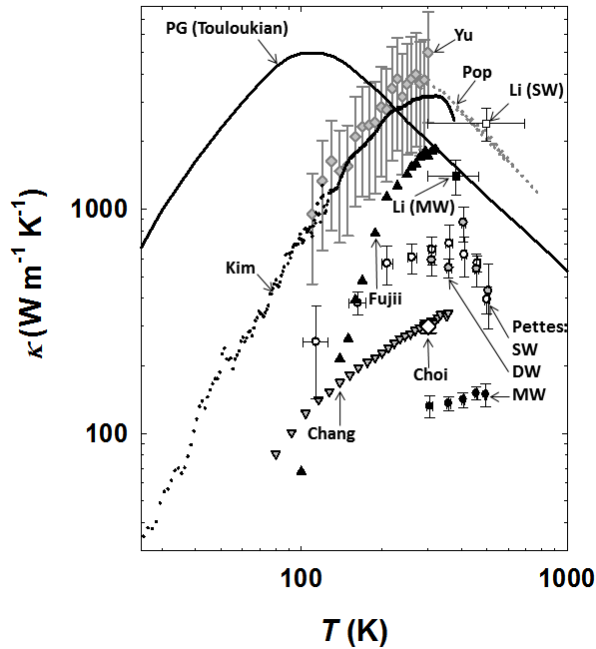

Figure 4. Experimental thermal conductivity values of CNTs reported by Kim *et al.* [60] for a ~14-nm-outer diameter and 2.5-μm-long MWCNT grown by arc discharge, by Fujii *et al.* [79] for a 16.1-nm-outer diameter, 4.9-nm-inner diameter, and 1.89-μm-long MWCNT grown by arc discharge, by Chang *et al.* [99] for a 33-nm-outer diameter MWCNT grown by arc discharge, by Choi *et al.* [82] for a ~20-nm-outer diameter, ~10-nm-inner diameter, and 1.4-μm-long MWCNT grown by plasma-enhanced CVD, by Yu *et al.* [98] for a 1–3 nm outer diameter and 2.76-μm-long SWCNT grown by thermal CVD, by Pop *et al.* [84] for a 1.7-nm-outer diameter and 2.6-μm-long SWCNT grown by thermal CVD, by Li *et al.* [94] for a 1.8-nm-diameter SWCNT and a and a 8.2-nm-diameter MWCNT grown by thermal CVD, by Pettes *et al.* [97] for a 1.5-nm-diameter, 2.03-μm-long SWCNT, 2.7-nm-outer diameter, 2.3-nm-inner diameter, and 4.02-μm-long DWCNT, and a 11.7-nm-outer diameter, 6.7-nm-inner diameter, 1.97 μm-long, 7-wall MWCNT, all grown by thermal CVD. Shown for comparison are the basal plane thermal conductivity recommended by Touloukian [112] for pyrolytic graphite.

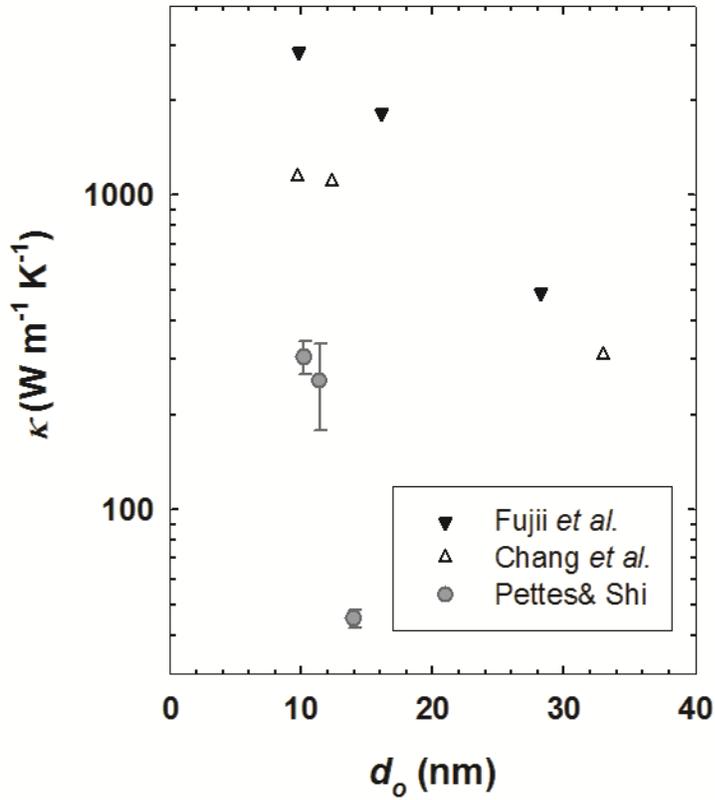

Figure 5. Thermal conductivity of MWCNTs of different diameters reported by Fujii *et al.* [79] for MWCNTs grown by arc discharge, by Chang *et al.* [99] for MWCNTs grown by arc discharge, and by Pettes and Shi [97] for MWCNTs grown by thermal CVD.

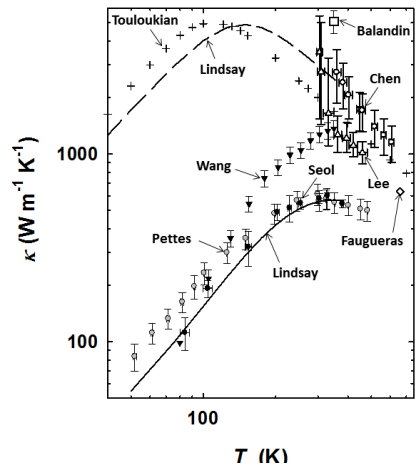

Figure 6. Experimental thermal conductivity values of graphene reported by Balandin *et al.* [105] for SLG samples exfoliated from pyrolytic graphite and suspended across a 2-5 μm wide trench, Faugeras *et al.* [109] for a SLG sample exfoliated from natural graphite and suspended over a 44 μm diameter hole, Chen *et al.* [111] for a SLG sample grown by CVD and suspended over a 9.7-μm-diameter hole, Lee *et al.* [113] for a SLG sample exfoliated from natural graphite and suspended over a 6-μm-diameter hole, Seol *et al.* [70] for a 9.5-μm-long, 2.4-μm-wide SLG sample exfoliated from natural graphite and supported on $SiO_2$, Wang *et al.* [136] for 5, 3, or 1-μm long three-layer graphene samples exfoliated from graphite and supported onto $SiO_2$, and Pettes *et al.* [71] for a 5-μm-long, 1.8-μm-wide suspended BLG sample exfoliated from natural graphite and with polymer residual. Shown for comparison are the basal plane thermal conductivity recommended by Touloukian [112] for pyrolytic graphite, the calculation results of Lindsay *et al*. [70] for 10-μm-long suspended (dashed line) and supported (solid line) SLG samples. For Raman measurements, the temperature is the temperature measured by the Raman laser beam while the substrate temperature is kept at room temperature.

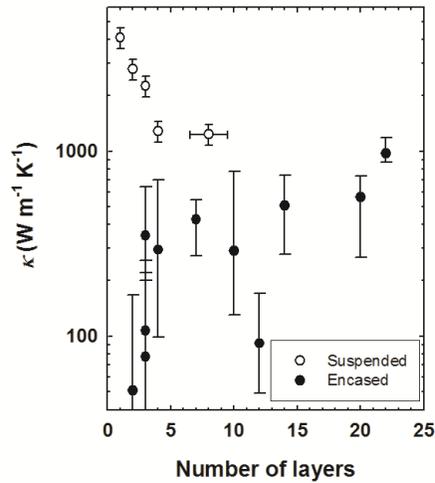

Figure 7. Thermal conductivity of graphene samples of different layer thickness reported by Ghosh *et al.* [138] for suspended graphene samples and by Jang *et al.* [137] for graphene samples encased in $SiO_2$. In the work of Ghosh *et al.* [138], experimental data obtained for 5-16 µm wide graphene samples suspended over 1-5 µm wide trenches were normalized with the use of a theoretical model to yield the reported data expected for 5-µm-wide suspended graphene samples.

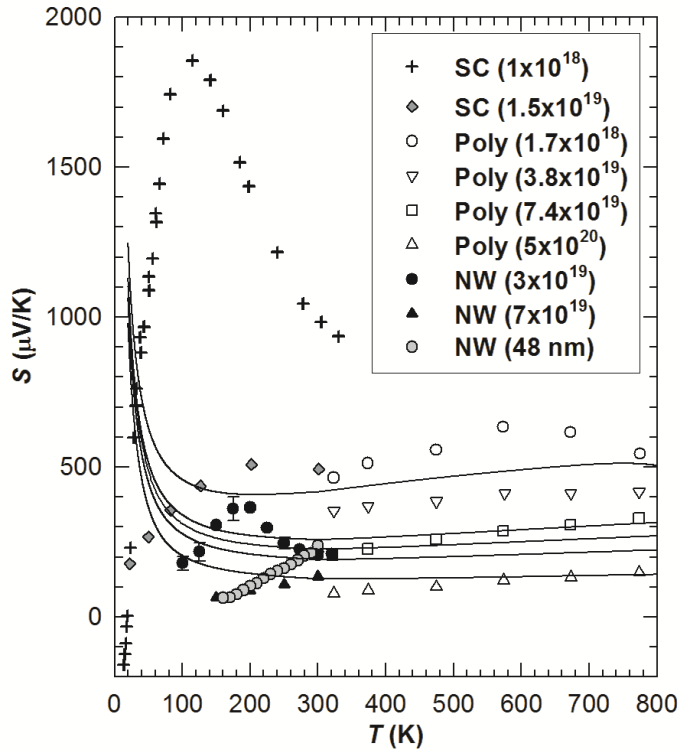

Figure 8. Measured Seebeck coefficient (symbols) of boron-doped Si reported by Geballe and Hull [151] for single-crystal (SC) samples of $1 \times 10^{18}$ and $1.5 \times 10^{19}$ cm$^{-3}$ acceptor concentrations, by Yamashita and Sadatomi [154] for polycrystalline (Poly) samples of $1.7 \times 10^{18}$, $3.8 \times 10^{19}$, $7.4 \times 10^{19}$ and $5 \times 10^{20}$ cm$^{-3}$ carrier concentration, by Boukai *et al.* [42] for a $20 \times 20$ nm$^2$ NW doped to $3 \times 10^{19}$ cm$^{-3}$ and a $10 \times 20$ nm$^2$ NW doped to $7 \times 10^{19}$ cm$^{-3}$, and by Hochbaum *et al.* [41] for a highly doped 48-nm-diameter NW. The lines are calculated $S_d$ results of a two-band model for boron doping concentration of $1 \times 10^{18}$, $1.5 \times 10^{19}$, $3 \times 10^{19}$, $7 \times 10^{19}$, and $5 \times 10^{20}$, respectively, arranged in the top to bottom order in the figure.

Graphic Submitted for Consideration as a Cover Art: A false color SEM image of a resistance thermometer device used for thermal measurement of supported SLG by Seol *et al.* [70].